\documentclass[journal=jpccck,manuscript=article,layout=twocolumn]{achemso}

\usepackage[version=3]{mhchem} % Formula subscripts using \ce{}
\usepackage{graphicx,epsfig,amsmath,amssymb}

\usepackage{color}
\usepackage{ulem}
\usepackage{comment}

\newcommand{\rem}[1]{\textcolor[rgb]{1,0,0}{\sout{}}}

\newcommand{\EQ}[3]{		%einzelne Gleichungen, 1:label, 2:gleichung, 3:satzzeichen
		\begin{equation}
		\label{#1}
		#2\;\;#3
		\end{equation}
		}
\newcommand{\ie}{i.\,e.~}	%analog
\newcommand{\etal}{\textit{et al.}~} %einfacheres et al.
\newcommand{\etalol}{\textit{et al.}} % et al. ohne Leerzeichen

\newcommand{\cytc}{cytochrome \textit{c}~}
\newcommand{\cytcol}{cytochrome \textit{c}}	% ohne Leerzeichen
\newcommand{\cytcc}{Cytochrome \textit{c}}	% ohne Leerzeichen
\newcommand{\tn}[1]{\textrm{#1}} % für normale Schrift in Formeln

\newcommand{\dC}{$^{\circ}$C}

\usepackage{color}
\title{pH-Dependent Selective Protein Adsorption into Mesoporous Silica}%

\author{Sebastian T. Moerz}%
\author{Patrick Huber}%
\email{patrick.huber@tuhh.de}

\affiliation{Institute of Materials Physics and Technology, Hamburg University of Technology (TUHH), D-21073 Hamburg-Harburg, Germany}%
\affiliation{Experimental Physics, Saarland University, D-66041 Saarbruecken, Germany}%

\email{patrick.huber@tuhh.de}

\begin{document}
\begin{tocentry}
\includegraphics[width=0.9\columnwidth]{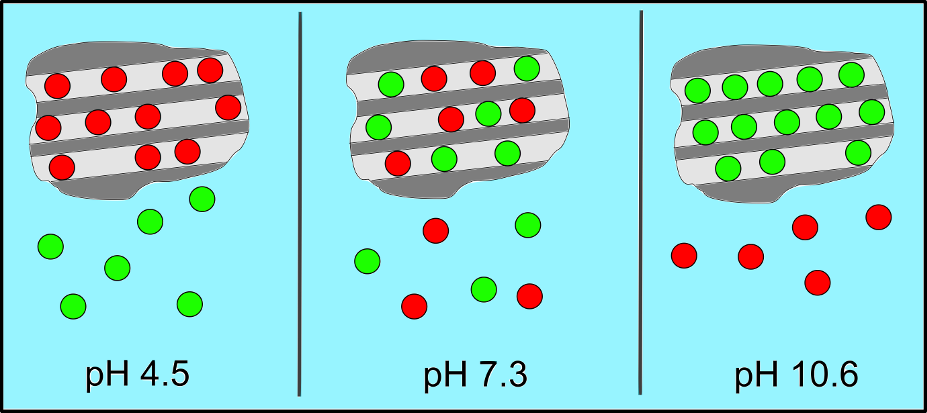}
%\epsfig{file=TOCC32SiliconCoating.eps, angle=0,width=0.5\columnwidth}
%\end{tocentry}
%\begin{tocentry}
%\includegraphics{TOCC32SiliconCoating.tif}
\end{tocentry}
%%%%%%%%%%%%%%%%%%%%%%%%%%%%%%%%%%%%%%%%%%%%%%%%%%%%%%%%%%%%%%%%%%%%%
%% Make a title, of course
%%%%%%%%%%%%%%%%%%%%%%%%%%%%%%%%%%%%%%%%%%%%%%%%%%%%%%%%%%%%%%%%%%%%%
\newpage
\begin{abstract}
The adsorption of lysozyme, \cytc and myoglobin, similar-sized globular proteins of approximately 1.5~nm radius, into the mesoporous silica material Santa Barbara Amorphous-15 (SBA-15) with 3.3~nm mean pore radius has been studied photometrically for aqueous solutions containing a single protein type and for binary protein mixtures. Distinct variations in the absolute and relative adsorption behavior are observed as a function of the solution's pH-value, and thus pore wall and protein charge. The proteins exhibit the strongest binding below their isoelectric points pI, which indicates the dominance of electrostatic interactions between charged amino acid residues and the -OH groups of the silica surface in the mesopore adsorption process. Moreover, we find for competitive adsorption in the restricted, tubular pore geometry that the protein type which shows the favoured binding to the pore wall can entirely suppress the adsorption of the species with lower binding affinity, even though the latter would adsorb quite well from a single component mixture devoid of the strongly binding protein. We suggest that this different physicochemical behavior along with the large specific surface and thus adsorption capability of mesoporous glasses can be exploited for separation of binary mixtures of proteins with distinct pI by adjusting the aqueous solution's pH.
\end{abstract}
%250 words abstract; 

%\section{Introduction}
% < 1000 words
\section{Introduction}
Over the last decades much attention has been payed to the adsorption and immobilization of charged, complex molecules at planar surfaces and in porous materials \cite{Netz1999, adiga2009, Schmitt2010, Felsovalyi2011, Lazzara2010, Bhattacharyya2010, McLoughlin2013, Steri2013, Meder2013,Zhou2013, Winkler2014, Kiesel2014, Retamal2014, Hildebrand2015, Huber2015, Carvalho2015, Meissner2015}. Especially nano- or more strictly spoken mesoporous materials and their interaction with biomolecules have been in the focus of intensive research, since they combine a high inner surface area with pore sizes large enough to comfortably host most proteins and enzymes \cite{adiga2009, Zhou2013}. The analogies between artificial nanopores and biological transport pores in biomembranes \cite{Zimmermann2011, Mahendran2013} and the existence of biomaterials consisting of proteins embedded in mesoporous matrixes\cite{Fratzl2007, Zlotnikov2014} has also stimulated fundamental studies on biomolecular adsorption, diffusion, and translocation processes in such geometric confinement \cite{Javidpour2008, Javidpour2009, Firnkes2010, Lee2012, Plesa2013, Mihovilovic2013}. Moreover mesoporous media promise a vast variety of applications in biochemical technologies, like enzymatic catalysis\cite{Hartmann_2005, Zhou2013, Fried2013}, protein crystallization \cite{Chayen2006, Meel2010} {and the fractionation of biological fluids like blood into their individual components}\cite{Yang2006}. New means of targeted drug delivery\cite{slowing_2007, adiga2009, Seker2012, Xue2012, Kurtulus2014, Argyo2014, Niedermayer2015} and drug design \cite{Graubner2014}, biosensors \cite{Janshoff1998, Kilian2009, Guan2011,Chen2011, Fan2012} and nanocomposite materials \cite{Boecking2012} can be envisaged. The optical transparency of mesoporous glasses along with the robustness of the fluidity of liquids at the nanoscale \cite{Eijkel2005, Gruener2009a, Gruener2015} allows one to employ them as versatile subunits in Lab-on-a-chip \cite{Dittrich2006} or more generally spoken microfluidic systems \cite{Stone2004, Hu2011, Yazdi2012, Bocquet2014}. Electrically conductive nanoporous solids, such as nanoporous gold or nanoporous carbon, additionally allow one to reversibly change the fluid/pore wall interaction by the application of electrical potentials and thus permit a versatile, external control of adsorption, guest-host interaction and fluid flow processes \cite{Xue2014}.

Especially the separation of a single protein type from a mixture containing a multitude of different biomolecules is of great importance for science as well as for technical applications. Three prominent ways to facilitate such a fractionation are dialysis, filtration and chromatography. While adsorption of proteins does not play a role in dialysis, it can be exploited to customise filtration applications\cite{causserand_2001} and {selective adsorption} is the fundamental principle behind chromatography.

An obvious way to separate a protein mixture is mere size-selectivity \cite{Sun2011,Qian2014,Yue2015}. When the pores of a filtering membrane are too small for one protein type to enter, but large enough for a second to pass, forcing the solution to flow or diffuse through the membrane will restrain the larger type \cite{Striemer2007, Uehara2009, Hu2011}. Successful size-selective separation of a binary protein mixture has been demonstrated by Katiyar \etalol\cite{katiyar_2006}. This study shows that the pores of the {host material} have to be slightly larger than the hydrodynamic radius of the protein to ensure adsorption. Yet mere size-exclusion is not the only factor determining the selective adsorption from multi-component solutions: Both the pore size and the protein-surface interaction play a crucial role {in} selective adsorption\cite{fujii_2006}. For example, electrostatic exclusion from the membrane is another means of filtration and can be used to separate two protein types of similar size but different net charge. A combination of protein adsorption and subsequent size-selective filtration was published by Causserand \etalol \cite{causserand_2001} in 2001.  

In this study, we use the mesoporous silica material Santa Barbara Amorphous-15 (SBA-15) to explore the competitive adsorption from binary mixtures of lysozyme, \cytc and myoglobin. These three {proteins} have similar hydrodynamic radii of approx. 1.5 nm as estimated from small-angle X-ray scattering experiments \cite{Segel1998, Zhou2013} and their comparable molecular mass. They fit comfortably into the SBA-15 pores, thus size-selectivity is unlikely to play any role. 

As shown in previous studies\cite{Zhou2013} the adsorption of native proteins in low ionic strength solutions is dominated by electrostatic interaction between charged amino acid residues and the -OH groups of the silica surface. This attractive interaction competes with the coulombic repulsion between equally charged protein molecules. At the isoelectric point, where the overall charge of the protein vanishes, this repulsion becomes negligible, the protein-silica interaction is dominated by van-der-Waals as well as polar interactions and a dense packing of molecules into the pores is observed\cite{sang_vinu_coppens_2011}. While similar in size, the three proteins used in this study differ in their respective isoelectric points. We demonstrate that this different physicochemical behavior results in a distinct adsorption behavior as a function of pH.

\section{Experimental}
\subsection{SBA-15 preparation and characterization}
The synthesis of hexagonally ordered mesoporous SBA-15 was first reported by Zhao \etal\cite{zhao1998}. The samples used in this thesis were prepared according to the following precedure: We mix 4~g of the tri-block co-polymer PEO$_{20}$-PPO$_{70}$-PEO$_{20}$ with 129.6~g water and 19.3~ml HCl (37\%). Due to its amphiphilic nature, the polymer forms an ordered phase of micellar structures when mixed with water. Vigorous stirring at 350 rpm for four hours is needed to ensure a homogenous emulsion. The mixture is kept in an oil bath at 55~\dC ~during this process. We then add 8.65~g tetraethylorthosilicate and stir the system for another 20 hours. We subsequently increase the temperature to 85~\dC ~and let the mixture rest for another 22 hours without stirring. During this time, the silicon from the TEOS leads to an accumulation of silica around the polymer micelles. These aggregates precipitate as a fine-grained powder. Calcination of the repeatedly rinsed powder at 500~\dC ~finally removes the polymer while preserving a negative of the micellar sturcture in the silica grains. The porous silica powder can now be used without further treatment or purification. 

The as-prepared SBA-15 was characterised by measuring volumetric nitrogen sorption isotherms in a custom-made, all-metal gas handling system. \rem{Again, a detailed discussion of the nitrogen sorption can be found in the study's first part. In short, we used} {The isotherms were analysed using} the mean-field model for capillary condensation proposed by Saam and Cole\cite{sc1974}\textsuperscript{,}\cite{sc1975}.\rem{to calculate the pore size distribution of SBA-15 sample. Careful analysis} {This} yields a bimodal pore size distribution consisting of a broad Gauss peak ($r_{\rm micro}=0.75$~nm with $\sigma_{\rm micro}=78\%$) corresponding to a pronounced microporosity\cite{imperorclerc2000} and a narrower peak ($r_{\rm meso}=3.3$~nm with $\sigma_{\rm meso}=6.5\%$) which {is}\rem{can be} attributed to \rem{approximately} cylindrical nanopores. \rem{that account for the bulk of the pore volume} A BET analysis of the sorption isotherms yields a specific inner surface of 613.9~m$^{2}$/g. Small angle x-ray diffraction reveals a hexagonal arrangement of these mesoporous channels with a lattice parameter $a_{\rm h}=10.71\pm0.08$~nm \cite{Hofmann2005, Zickler2006}. Since the molecules studied here are far too large to enter the micropores, the microporous fraction of the samples was neglected in this study. 

We have not performed pH-dependent measurements of the $\zeta$-potential of SBA-15 and thus of the charging of the mesoporous particles. Given its importance for adsorption processes in water \cite{Rosenholm2008}, there is, however, sizeable literature available with regard to the pH-dependence of the $\zeta$-potential of silica surfaces \cite{Kirby2004} and of SBA-15 in particular \cite{Rosenholm2008}. For the considerations following below, it is interesting to note that the isoelectric point of SBA-15 has been reported as pH\,3.8\cite{Essa_2007}.

\subsection{Proteins}

Bovine heart cytochrome \textit{c} (purity $\geq$95\%, catalog number C2037), equine skeletal muscle myoglobin (purity 95-100\%, catalog number M0630) and chicken egg white lysozyme (purity $\geq$90\%, catalog number L6876) were purchased from Sigma Aldrich and used without further treatment or purification. According to the distributor, the proteins' masses are 12.3~kDa, 17.6~kDa and 14.3~kDa and the isoelectric points pH 10.0-10.5, pH 7.3 and pH 11.35, respectively. 

\subsection{Buffer solutions}

10mM buffers were prepared according to the following recipes: pH 3.0 (116.37~g of 0.01~M citric acid and 4.0~ml of 0.01~M trisodium citrate), pH 3.8 (200.02~g of 0.01~M acetic acid and 15~ml of 0.01~M sodium acetate), pH 4.5 (100.23~g of 0.01~M trisodium citrate and 120~ml of 0.01~M citric acid), pH 6 (100.14~g of 0.01~M monopotassium phospate and 32.5~ml of 0.01~M sodium hydroxide), pH 7.3 (50.40~g of 0.01~M sodium hydroxide and 41.7~ml of 0.05~M monopotassium phospate), pH 8.5 (83~ml of 0.01~M sodium tetraborate and 13.1~ml of 0.01~M hydrochloric acid) and pH 10.6 (100~g of 0.01~M sodium bicarbonate and 71~ml of 0.01~M sodium hydroxide). PH values were calibrated using a \textit{Mettler Toledo Seven Easy} pH meter equipped with an \textit{InLab Expert Pro} electrode.

%\EPSB{referenzspektren2.png}{.40\textwidth}{ref_spec}{Reference spectra of the three different proteins at different pH, shifted with a vertical offset for better visibility. Top : Lysozyme. Middle: \cytcc. Bottom: Myoglobin.}{} 

\begin{figure}[tbp]
\includegraphics[width=0.9\columnwidth]{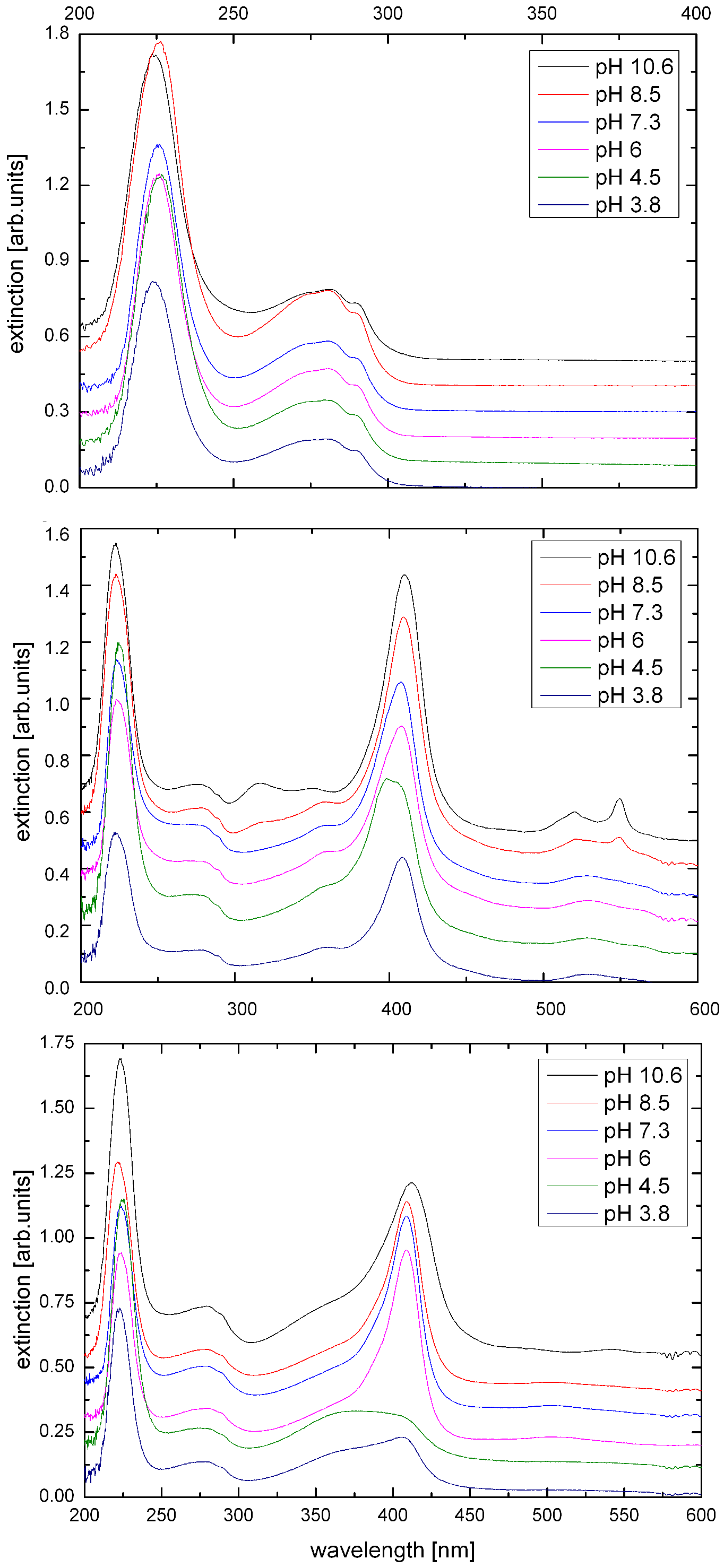}
\caption{Reference spectra of the three different proteins at different pH, shifted with a vertical offset for better visibility. Top : Lysozyme. Middle: \cytcc. Bottom: Myoglobin.} \label{ref_spec}
\end{figure}

\begin{figure}[tbp]
\includegraphics[width=1\columnwidth]{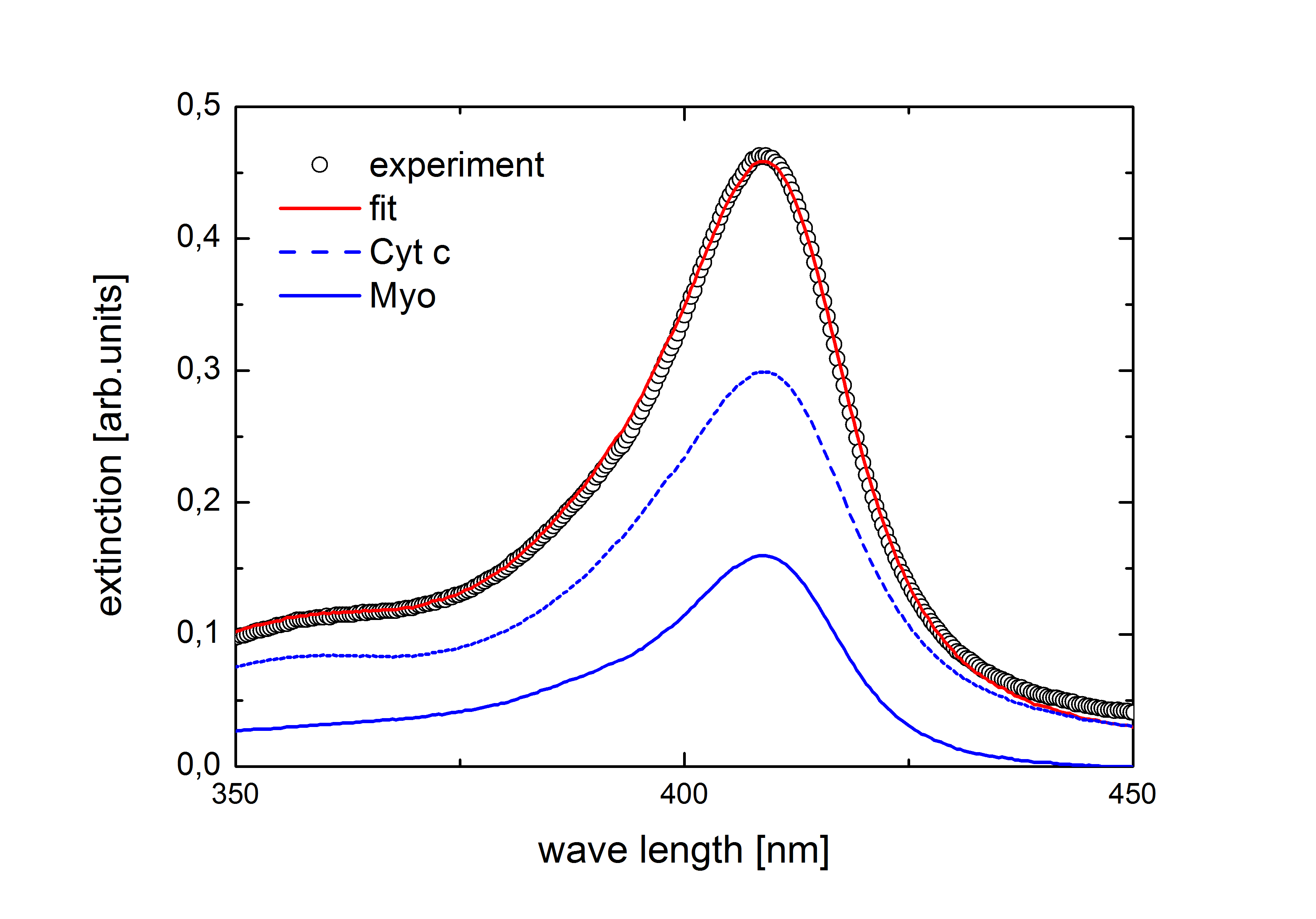}
\caption{Linear deconvolution of a UV-Vis spectrum of the supernate after competitive adsorption of \cytc and myoglobin at pH=6. Plotted are the experimental data along with the reference spectra of the single proteins weighted according to their contribution and the corresponding sum spectrum (fit). } \label{fig:UVVISCMpH6}
\end{figure}

\subsection{Protein adsorption experiments}

Single protein solutions containing 2~g/L of a single protein and binary mixtures containing two different proteins with 1~g/L each were prepared with the buffers described above. Small amounts of SBA-15 ($4.0-9.0~\tn{mg}$, measured with a \textit{Sartorius type 1801} analytical balance) were mixed with appropriate amounts of protein solution (200~$\mu$l per mg of SBA-15) and transferred to \textit{Eppendorf Safe-Lock} tubes. {The silica powder was dispersed by ultrasonification for 10 minutes.} The tubes were equipped with small stirring bars and transferred to a water bath kept at 31\dC. Both the bath and the samples were stirred at 350 rpm using an IKA \textit{RCT basic safety control} magnetic stirrer. \rem{Ultrasonification for 10 minutes ensured a fine dispersion of the silica grains.} The samples were kept in the bath for five days, which was found to be sufficient to reach equilibrium\cite{moerz_2014}.\textsuperscript{,}\cite{zhang_2007}\textsuperscript{,}\cite{Vinu_2004}  The protein concentrations in the supernate were determined photometrically. UV-Vis absorbance spectra were recorded with an \textit{Ocean Optics USB2000-UV-VIS-ES} spectrometer and a \textit{DT-MINI-2-GS UV-VIS-NIR} light source. The light source uses a deuterium and a halogen bulb to provide a continuos spectrum ranging from 200 to 850~nm. The spectrometer and the light source were connected to a cuvette holder via appropriate fiber optics. Disposable semi-micro UV-cuvettes (\textit{Brand}, cat. no. 7591 50.) held the sample aliquots during the photometrical analysis. Note that we found no hints for protein degradation in UV-Vis absorbance spectra over time periods typical for our experiments (120 hours). This is maybe not too surprising given the small, globular character of the molecules studied here. The individual supernate concentrations after the adsorption were calculated by performing a multiple linear regression of the supernate spectrum $\alpha_{\rm S} (\lambda)$

\EQ{mult_lin}{\alpha_{\rm S} (\lambda) = c_{\rm 1} \cdot \alpha_{\rm A} (\lambda) + c_{\rm 2} \cdot \alpha_{\rm B} (\lambda)}{}

using the respective reference spectra $\alpha_{\rm A}$ and $\alpha_{\rm B}$ of the mixtures' two components and their concentrations $c_{\rm 1}$ and $c_{\rm 2}$ in units of gram per litre. The reference spectra of the three different proteins recorded at different pH values are shown in Figure \ref{ref_spec}. In Fig. \ref{fig:UVVISCMpH6} a deconvolution of a supernate spectrum is exemplified. Finally, the amount of adsorbed protein per gram of the mesoporous host material was calculated from the difference of the initial and final protein concentrations. 
\section{Results and discussion}

\rem{Lysozyme is known to be a very rigid and stable molecule. This is well-reflected in the data which show no apparent sign of structural changes except for the highest pH value.}{As shown in the top panel of Figure \ref{ref_spec} the spectrum of lysozyme was independent of the pH value except for the most alkaline buffer.} The \cytc data are shown in the middle panel. Above pH~8.5 the distinct peaks in the Q band region between 500 and 550~nm indicate a change in the protein's oxidation state from ferric to ferrous\cite{oellerich_2002}. However, this transition should not affect our experiments, since the oxidation state has no significant influence on the adsorption behavior of folded cytochrome \textit{c}\cite{kraning_2007}. Cytochrome \textit{c} appears to be stable for all pH values. The broadening and blue-shift of the Soret band of the pH~4.5 sample indicates a transition of the central iron ion from a low-spin to a high-spin state and is not a sign of a larger structural change\cite{oellerich_2002}. The Soret peak of myoglobin exhibits a broadening for both high and low pH values. Nevertheless, we used these spectra to calculate the supernate concentration according to equation \ref{mult_lin}. 
To estimate the maximum pore loading of the different proteins at the examined pH values, we mixed SBA-15 samples with solutions containing 2~g/L of the respective protein. Based on our previous findings\cite{moerz_2014} we assume that this initial concentration is sufficient to estimate the saturation value of the Langmuir-type isotherms. The adsorption was carried out as described above.

\begin{figure}[tbp]
\includegraphics[width=1\columnwidth]{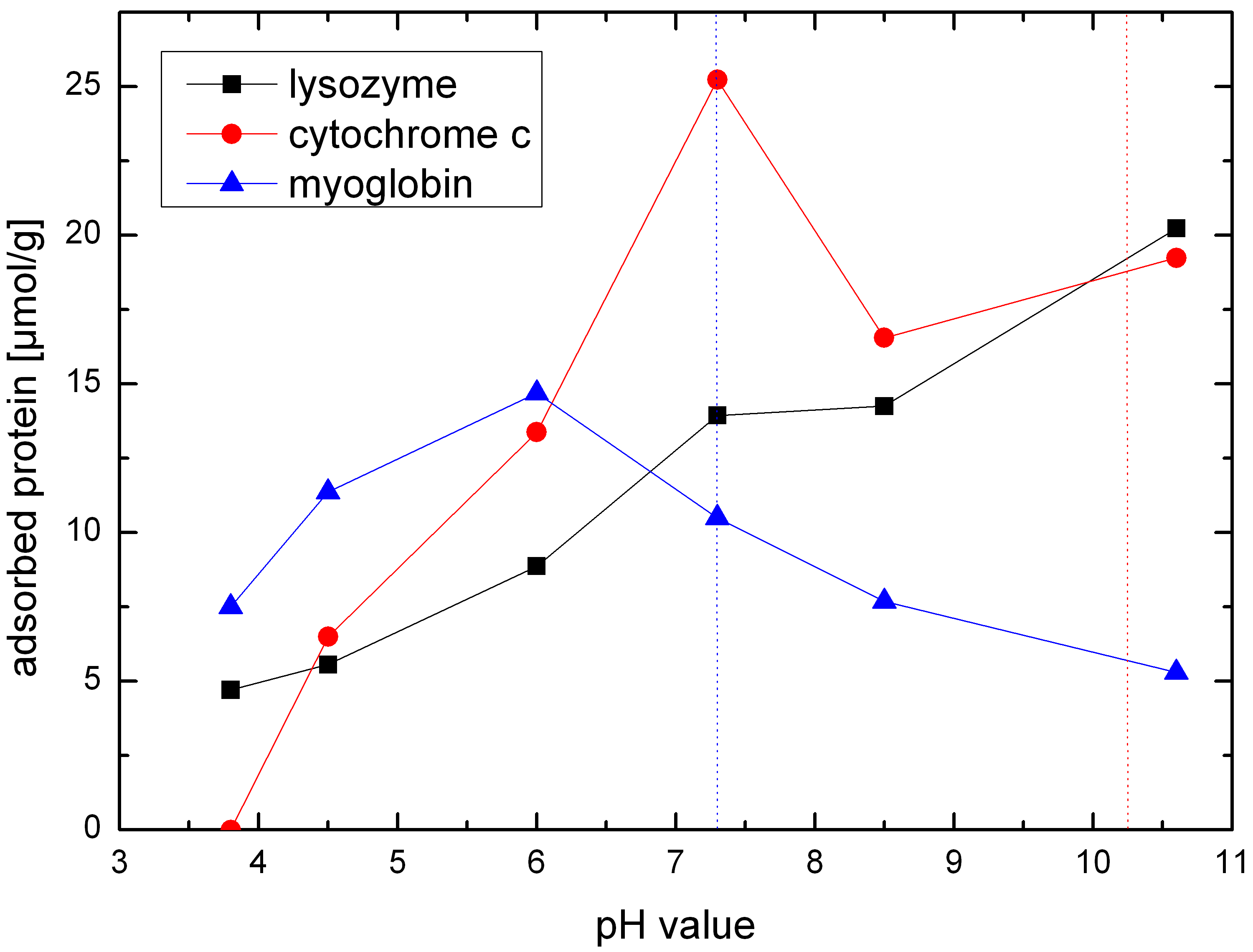}
\caption{Maximum pore loading of SBA-15 for lysozyme (black squares), \cytc (red circles) and myoglobin (blue triangles) for different pH values in 10~mM buffer solutions.} \label{selec1}
\end{figure}

In Figure \ref{selec1} we show the measured maximum pore loadings in $\mu$mol per gram of SBA-15 as a function of the buffer pH. The black squares represent the lysozyme data, while the red circles and blue triangles denote \cytc and myoglobin, respectively. Neither protein showed any adsorption at pH~3 where both the surface and the protein have the same charge sign\cite{Essa_2007}. The pH~3 buffer is therefore neglected for the following measurements. The maximum of lysozyme adsorption is beyond the scope of the pH values investigated, as is its isoelectric point. Both \cytc and myoglobin exhibit clear maxima considerably below their isoelectric points, which are indicated by the vertical blue dotted line (myoglobin) and the red dotted line (\cytcol). Myoglobin still shows a non-neglectable pore loading even in the most alkaline buffers \rem{used here, even though one would expect}{where} strong electrostatic repulsion between the protein and the surface {is expected}\rem{which are both negatively charged in these buffers}. {The observed adsorption can be explained by positively charged patches on the protein's surface which are present despite the molecule's negative overall charge. Interfacial charge regulation may also contribute to an attractive interaction even above the pI\cite{Hartvig2011}.} 
%Note that this behaviour contradicts the findings of Katiyar \etalol\cite{katiyar_2005} who found a strong decrease of myoglobin pore loading in SBA-15 above the isoelectric point. \rem{The bigger size of the myoglobin molecule causes a lower maximum value for the pore loading when compared to \cytcol. The ratio of the maximum pore loadings is approximately 0.6, close to the ratio of the molecule masses $\frac{12.3~\tn{kDa}}{17.6~\tn{kDa}} \approx$ 0.7.}

With respect to the pH-dependent adsorption of the proteins the stability of mesoporous SBA-15 in aqueous solution should be discussed. Pham \etal \cite{Pham2012} performed a thorough study of the dissolution of SBA-15 as a function of pH. They find that this matrix is increasingly unstable for pH values larger than 7. E.g. in their column experiments, more than 45\% of the initial SBA-15 dissolved within 48 hours, when a pH 8.5 solution flowed through the column. In our experiments, where the mesoporous silica is exposed to the solution for more than 120 hours, we found no hints for a dissolution of the matrix in the alkaline regime. This may be related to the fact that, in contrast to the experiments by Pham \etal, we did not permanently exchange the solution. Moreover, the protein adsorption possibly reduces the silica degradation process in the alkaline regime. 
 
%ption species artefecact apparent reduction. betweWe assume that the decrease in protein adsorption for all proteins could be explainable by this effect. In particular, this ''apparent" adsorption reduction will result in a drop of the specific adsorption before reaching of the isoelectric point of the corresponding protein, as observed in our experiments. By the same token, the counter-intuitive observation of a maximum adsorption of the proteins at significant lower values can be considered as an hint of the SBA-15 dissolution in the basic state.

\begin{figure}[tbp]
\includegraphics[width=1\columnwidth]{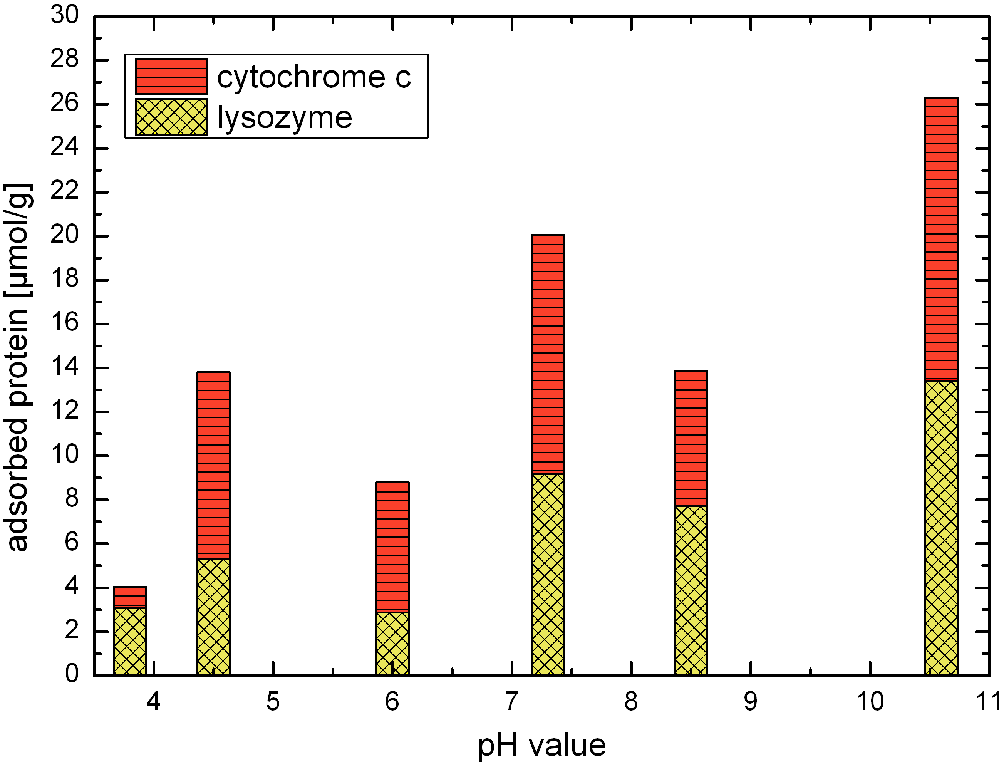}
\caption{Proteins adsorbed to SBA-15 from a binary mixture of \cytc and lysozyme.} \label{selec2}
\end{figure}

The bar graph in Figure \ref{selec2} shows the resulting pore loadings from competitive adsorption from a binary mixture of \cytc and lysozyme. The overall height of the bar denotes the total amount of protein bound to the SBA-15 sample, while the differently dashed segments represent the shares that can be attributed to the different proteins involved. In this graph, the horizontally dashed segments represent the bound \cytcol. The checkered bar represents adsorbed lysozyme. Both proteins attribute to roughly one half of the total pore loading at pH~10.6, while \cytc adsorption is slightly prevalent at lower pH values. The buffer with pH~3.8 marks an exception, where lysozyme amounts for most of the pore filling. Both proteins adsorb in mutual presence for all pH values and we do not observe any effective fractionation of the binary mixture. This is hardly surprising since both molecules are of similar size and \rem{rather comparable in their}isoelectric points.

\begin{figure}[tbp]
\includegraphics[width=1\columnwidth]{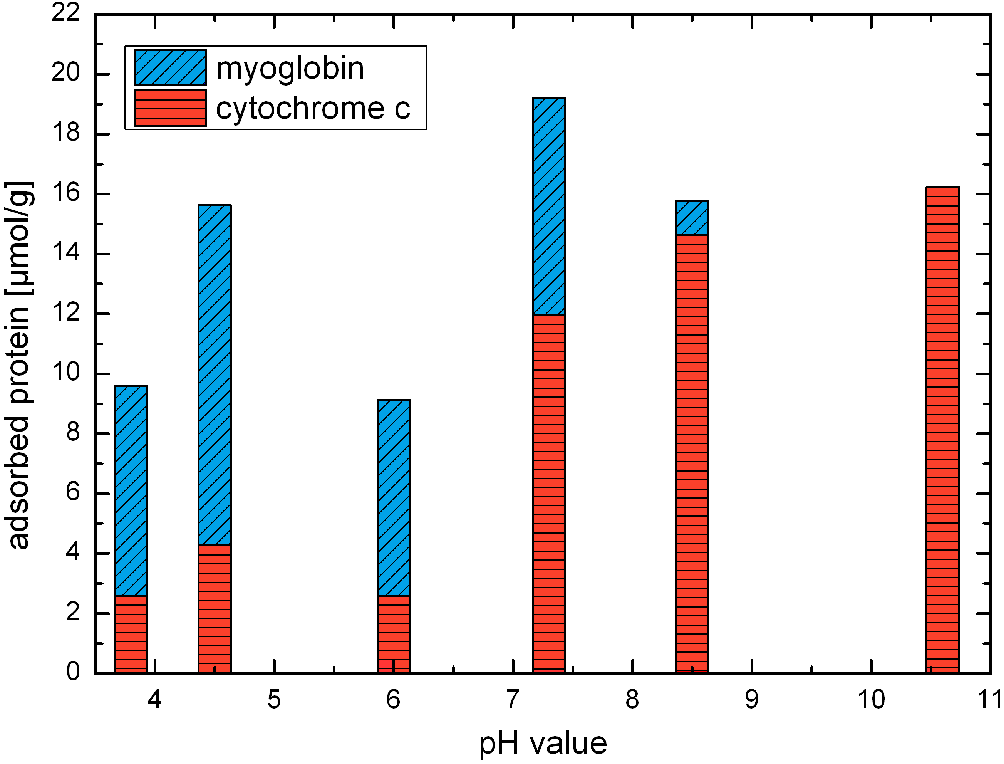}
\caption{Proteins adsorbed to SBA-15 from a binary mixture of myoglobin and \cytc.} \label{selec4}
\end{figure}

We observe a different behavior for a mixture containing \cytc and myoglobin. As shown in Figure \ref{selec4}, there is a prevalence of myoglobin binding for pH~3.8, 4.5 and 6.0: {The strong positive charge of the \cytc leads to high protein-protein repulsion. The adsorption of \cytc is therefore less favorable than the adsorption of myoglobin, which is much closer to its pI. This trend is reversed above the myoglobin's isoelectric point were the overall charge of the protein becomes negative \ie, for all buffers with a pH of 7.3 and higher. The overall charge now leads to electrostatic repulsion between the silica and the myoglobin and consequently \cytc adsorption is favoured. Interestingly, myoglobin can still adsorb at high pH values from single protein solutions, as shown in Figure \ref{selec1}, but gets completely displaced at pH~10.6 when competing with \cytc. A closer look at the data reveals that not only is there no myoglobin bound to the silica particles, the observed pore loading corresponds to the entire amount of \cytc present in the initial solution, rendering the supernate devoid of this protein. This is possible because of the large specific surface and thus sorption capability of the mesoporous medium employed.}

\rem{This region is far from the cytochrome's isoelectric point and we expect to find a rather low binding affinity. Myoglobin, on the other hand, is much closer to its pI. It experiences a higher affinity to the surface than the \cytc and shows a much higher adsorption. This trend is inverted above the myoglobin's pI. While the residual interaction is still high enough to cause adsorption from a single component solution, the myoglobin molecules now bear a negative overall charge which reduces their binding affinity to the silica. It now loses the competition for adsorption sites to the still positively charged \cytcol. A small amount of myoglobin still binds at pH~8.5, but at pH~10.6 \cytc completely displaces its competitor. A closer look at the data reveals that not only is their no myoglobin bound to the silica particles, the observed pore loading corresponds to the entire amount of \cytc present in the initial solution, rendering the supernate devoid of this protein. An effective protein separation should be feasible at this pH value.}

\begin{figure}[tbp]
\includegraphics[width=1\columnwidth]{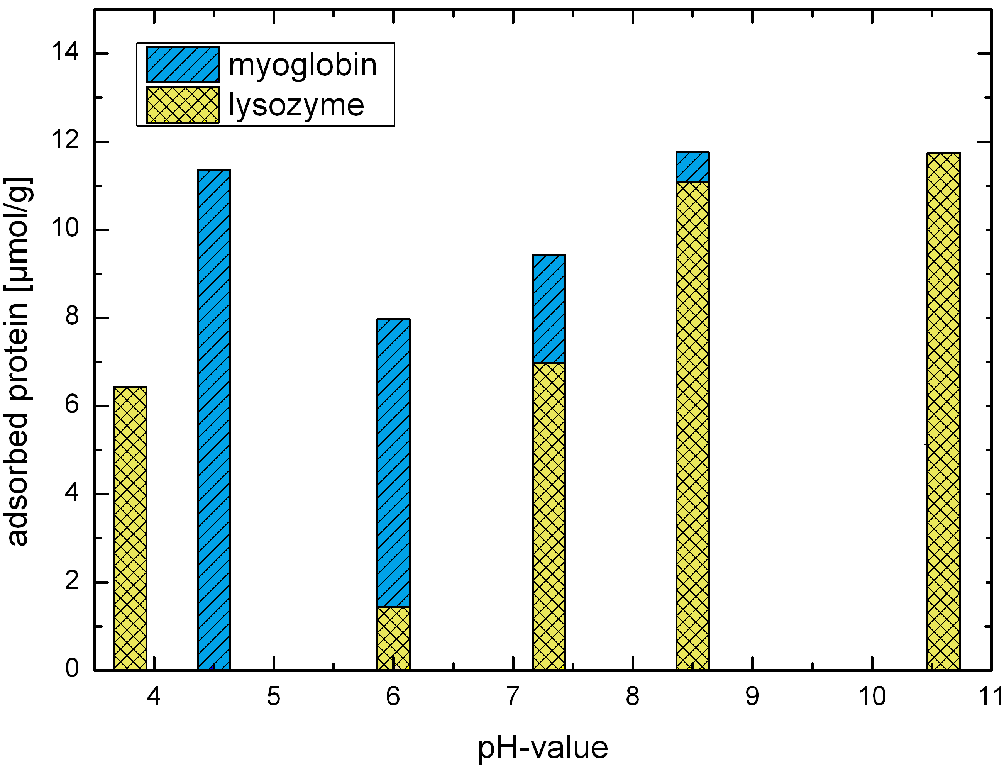}
\caption{Proteins adsorbed to SBA-15 from a binary mixture of myoglobin and lysozyme.} \label{selec3}
\end{figure}

%\EPSB{selec3_col.png}{.6\textwidth}{selec3}{}{Simultaneous adsorption of myoglobin and lysozyme} 

This becomes even more pronounced when comparing myoglobin with lysozyme. The data for their competitive adsorption are shown in Figure \ref{selec3}. We find exclusive binding of lysozyme at pH~10.6 with more than 85\% of the entire lysozyme immobilised in the silica pores and exclusive and complete binding of myoglobin at pH~4.5. The buffers at intermediate pH values show a smooth transition between these extremes. Surprisingly, we find a complete inversion of this behavior at pH~3.8. About 45\% of the lysozyme adsorbs to the SBA-15 while myoglobin is again completely displaced.

\rem{The data highlight the paramount importance of the protein's isoelectric point and thus the electrostatic interaction for adsorption of native, folded proteins at relatively low ionic strength, where the Debye length is on the same scale as the protein diameter\cite{pasche_2005}. Adsorption is strongly favoured close to the point where the protein's net charge vanishes. So strongly, in fact, that under certain circumstances the uncharged type can fully displace its charged competitors, even though they would adsorb quite well in the absence of the neutral protein. The sudden displacement of myoglobin by lysozyme at pH~3.8}

{This} is somewhat puzzling: The electroneutrality of the silica at these chemical conditions\cite{Essa_2007} readily explains a smaller affinity of the myoglobin to the surface. This should, however, also apply for lysozyme whose strong positive charge should cause a considerable repulsion between the molecules and thus inhibit any considerable adsorption. It seems plausible to assume that local \rem{alterations}{changes} of the pH near the silica surface\cite{Hartvig2011} might \rem{render}{alter} the {local} environment inside the pores \rem{so acidic that the myoglobin structure becomes unstable}{in a way that the proteins are no longer stable}, resulting in unpredictable changes of the adsorption behavior. Yet, a sound interpretation of this behavior is clearly beyond the scope of these experiments and complementary measurements in this pH-range are necessary in order to explore in detail the repeatability of this peculiar behavior.

\section{Conclusions}

We studied the competitive adsorption of lysozyme, myoglobin and \cytc to the mesoporous silica SBA-15 for a range of different pH values. Protein adsorption appears to be favoured in the vicinity of the protein's respective isoelectric point. Myoglobin and lysozyme, which differ strongly in their isoelectric points, can be effectively separated from a binary mixture if the buffer pH favours the adsorption of only one of the components. The type \rem{with the higher binding affinity adsorbs}{which shows the favoured binding} to the silica \rem{and}{can} replace the one with {the} lower binding affinity, even though the latter would adsorb quite \rem{considerable}{well} from a single component mixture devoid of the strongly binding protein. This separation was not observed for \cytc and lysozyme, which are very similar in their pIs.

The observation that the bulk solution can be completely devoid of the pore-preferred protein, whereas the mesoporous medium is for certain pH-values and protein mixtures exclusively filled with the preferred species highlights that the large adsorption capabilities along with the pH-dependent adsorption selectivity might allow for the design of customized devices for protein separation in binary mixtures using the nowadays readily available bulk mesoporous media. Note, however, that the sizeable dissolution rate of mesoporous silica in the alkaline regime discussed above may be a high technological hurdle. Possibly, it can be surpassed by surface-grafting and thus by a stabilization of the silica walls against alkaline dissolution. Albeit this will clearly necessitate additional studies on competitive adsorption and on SBA-15 stability in aqueous solutions. 

%We conclude that while other mechanisms, like counterion release and Van-der-Waals forces, certainly play a non-neglectable role,\cite{moerz_2014} {this study highlights the paramount importance of the protein's isoelectric point and thus the electrostatic interaction for adsorption of native, folded proteins at relatively low ionic strength, where the Debye length is on the same scale as the protein diameter\cite{pasche_2005}.}\rem{the pH-dependent separation of proteins with different isoelectric points clearly highlights the paramount importance of the electrostatic interaction for the adsorption of native proteins in low ionic strength solutions.} 

While the importance of the electrostatic interaction for protein adsorption is widely known and generally accepted \cite{Zhou2013}, this study is to the best of our knowledge the first one to demonstrate exclusive adsorption from a binary mixture rather than adsorption of a mixed layer. We anticipate that this high adsorption selectivity of the mesoporous medium in comparison to planar surfaces is also related to the tubular, confined sorption geometry. It makes it difficult for the adsorbing proteins to avoid interacting with each other, whereas in a more conventional planar geometry, they might be able to still adsorb at nearby sites as they are not so intimately contained and forced to interact. 

Therefore, we hope that our experimental findings will stimulate model calculations and computer simulation studies of protein adsorption in tubular geometries, similarly as this has been done recently for polyelectrolyte adsorption in charged cylindrical pores \cite{Carvalho2015} or for planar surfaces, that is in semi-infinite confined geometry \cite{Schmitt2010, Szott2011, Hildebrand2015}. 

In future experimental studies, it would be interesting to examine the pore size-dependence of pH-dependent competitive protein adsorption as well as the influence of distinct pore-size hierarchies as they are available nowadays for porous silica \cite{Inayat2013}. Similarly as observed for adsorption \cite{Huber1999} and transport \cite{Gruener2011, Kriel2014} of simple fluids, for molecular self-diffusion \cite{Kusmin2010, Hofmann2012, Valiullin2009} and for wetting of binary fluids \cite{Woywod2003}, one can expect a strong influence of the characteristic confinement size on competitive adsorption processes of charged macromolecular systems \cite{Carvalho2015}. Another interesting experimental parameter to modify is the surface chemistry, either by changing the porous host material or by pore wall grafting. In particular, the study of mesoporous carbon materials \cite{Presser2011, Liu2014} with positively or negatively charged surface groups as well as the exploration of mesoporous alumina \cite{Lu2012, El-Safty2013} could be very interesting, because of the higher pH-stability of these materials in comparison to their silica counterparts.

\section{Acknowledgments}
We acknowledge stimulating and helpful discussions with Stefan Bommer, Tihamer Geyer, Volkhard Helms, Martin Jung, and Richard Zimmermann (Saarland University). This work has been supported by the German Research Foundation (DFG) within the graduate school 1276, `Structure formation and transport in complex systems' (Saarbr\"{u}cken) and through the collaborative research initiative SFB 986 'Tailor-Made Multi-Scale Materials Systems  M3' (Hamburg) within the research area B.
\providecommand{\latin}[1]{#1}
\providecommand*\mcitethebibliography{\thebibliography}
\csname @ifundefined\endcsname{endmcitethebibliography}
  {\let\endmcitethebibliography\endthebibliography}{}

\end{document}